\documentclass[pra,groupedaddress,showpacs,twocolumn,nofootinbib]{revtex4-1}
\usepackage{amsmath}
\usepackage{amssymb}
\usepackage{amsthm}
\usepackage{graphicx}
\usepackage{epsfig}
\usepackage{epstopdf}

\begin{document}

\title{Entangling two oscillators with arbitrary asymmetric initial states}

\author{Chui-Ping Yang$^{1}$, Qi-Ping Su$^{1}$, Shi-Biao Zheng$^{2}$}
\author{Franco Nori$^{3,4}$}\email{fnori@riken.jp}
\author{Siyuan Han$^{5}$}

\address{$^1$Department of Physics, Hangzhou Normal University, Hangzhou, Zhejiang 310036, China}
\address{$^2$Department of Physics, Fuzhou University, Fuzhou 350002, China}
\address{$^3$CEMS, RIKEN, Saitama 351-0198, Japan}
\address{$^4$Department of Physics, University of Michigan, Ann Arbor, Michigan 48109-1040, USA}
\address{$^5$Department of Physics and Astronomy, University of Kansas, Lawrence, Kansas 66045, USA}

\date{\today}

\begin{abstract}
A Hamiltonian is presented, which can be used to convert any asymmetric
state $|\varphi \rangle _{a}|\phi \rangle _{b}$ of two oscillators $a$ and $%
b $ into an entangled state. Furthermore, with this Hamiltonian and local
operations only, two oscillators, initially in any asymmetric initial
states, can be entangled with a third oscillator. The prepared entangled
states can be engineered with an arbitrary degree of entanglement. A
discussion on the realization of this Hamiltonian is given. Numerical
simulations show that, with current circuit QED technology, it is feasible
to generate high-fidelity entangled states of two microwave optical fields,
such as entangled coherent states, entangled squeezed states, entangled
coherent-squeezed states, and entangled cat states. Our finding opens a new
avenue for creating not only two-color or three-color entanglement of light
but also wave-like or particle-like entanglement or novel wave-like and
particle-like hybrid entanglement.
\end{abstract}

\maketitle
\date{\today }

\textit{Introduction.} Entangled states of light are a fundamental resource
for many quantum information tasks [1-8]. Over the last two decades, much
attention has been devoted to the generation of entangled states of light.
In the regime of discrete variables, entanglement of up to eight photons has
been experimentally demonstrated via linear optical devices [9,10]. In the
regime of continuous variables, EPR states of light have been experimentally
generated from two independent squeezed fields [11,12], two independent
coherent fields [13], or a single squeezed light source [14]; two- or
three-color entangled states of light have been experimentally prepared by
means of non-degenerate optical parametric oscillators [15-17]. Recently,
hybrid entanglement between particle-like and wave-like optical qubits or
between quantum and classical states of light [18,19] has also been
demonstrated in experiments, which has drawn increasing attention because
hybrid entanglement of light is a key resource in establishing hybrid
quantum networks and connecting quantum processors with different encoding
qubits. Moreover, a large number of theoretical proposals have been
presented for generating \textit{particular types} of entangled states of
light or optical fields in various physical systems [20-33].

In this letter, we propose a Hamiltonian, which can be used to convert any
asymmetric state $\left\vert \varphi \right\rangle _{a}\left\vert \phi
\right\rangle _{b}$\ of two oscillators $a$\ and $b$\ into an entangled
state $\alpha \left\vert \varphi \right\rangle _{a}\left\vert \phi
\right\rangle _{b}\pm \beta \left\vert \phi \right\rangle _{a}\left\vert
\varphi \right\rangle _{b}$. Here the term asymmetric state refers to the
product state $\left\vert \varphi \right\rangle _{a}\left\vert \phi
\right\rangle _{b},$\ with $\left\vert \varphi \right\rangle \neq \left\vert
\phi \right\rangle $.\ The procedure consists of a single operation and a
posterior measurement on the states of the qudit coupler that is used to
couple the oscillators. Furthermore, by combining this Hamiltonian with
additional local operations, two oscillators $a$\ and $b$\ initially in any
asymmetric state $\left\vert \varphi \right\rangle _{a}\left\vert \phi
\right\rangle _{b}$\ and a third oscillator in the vacuum state $\left\vert
0\right\rangle _{c}$\ can be converted to a tripartite entangled state $%
\alpha \left\vert \varphi \right\rangle _{a}\left\vert \phi \right\rangle
_{b}\left\vert 0\right\rangle _{c}+\beta \left\vert \phi \right\rangle
_{a}\left\vert \varphi \right\rangle _{b}\left\vert 1\right\rangle _{c}$
with no measurement required. Hereafter, we call them the bipartite and tripartite protocols respectively.
In both cases, the degree of entanglement,
determined by the two coefficients $\alpha $\ and $\beta $, is adjustable by
controlling the initial state of the qudit coupler. The prepared two- or
three-oscillator entangled states can be two-color or three-color entangled
states when each oscillator has a different frequency. More importantly, the
light fields involved can be wave-like entangled states, particle-like
entangled states, or wave-like and particle-like hybrid entangled states,
depending on whether the states $\left\vert \varphi \right\rangle $ and $%
\left\vert \phi \right\rangle $ are both wave-like states (e.g., coherent
states, squeezed states, and cat states), particle-like states (e.g., Fock
states), or one wave-like and the other particle-like states (e.g., coherent
states and Fock states).

In contrast to previous works [11-33] aimed at generating particular types
of entangled states, this work provides a \textit{general} method for generating
various two- or three-oscillator entangled states.
Moreover, independent of the nature of the two non-identical states $%
\left\vert \varphi \right\rangle $ and $\left\vert \phi \right\rangle $, the
bipartite protocol requires post-selection by measurement while the tripartite protocol does not. So
they are not the \textquotedblleft same\textquotedblright . We note
that this proposal can be applied to create a set of interesting two-oscillator
entangled states, such as: (i) entangled \textit{wave-like} coherent states $%
\left\vert \alpha \right\rangle _{a}\left\vert -\alpha \right\rangle _{b}\pm
\left\vert -\alpha \right\rangle _{a}\left\vert \alpha \right\rangle _{b},$
(ii) entangled \textit{wave-like} squeezed states $\left\vert \xi
\right\rangle _{a}\left\vert -\xi \right\rangle _{b}\pm \left\vert -\xi
\right\rangle _{a}\left\vert \xi \right\rangle _{b},$ (iii) entangled
\textit{wave-like} cat states $\left\vert \mathrm{cat}\right\rangle
_{a}\left\vert \overline{\mathrm{cat}}\right\rangle _{b}\pm \left\vert
\overline{\mathrm{cat}}\right\rangle _{a}\left\vert \mathrm{cat}%
\right\rangle _{b}$ with cat states $\left\vert \mathrm{cat}\right\rangle
=\left\vert \alpha \right\rangle +\left\vert -\alpha \right\rangle $ and $%
\left\vert \overline{\mathrm{cat}}\right\rangle =\left\vert \alpha
\right\rangle -\left\vert -\alpha \right\rangle $; (iv) entangled \textit{%
wave-particle-like} coherent-Fock states $\left\vert \alpha \right\rangle
_{a}\left\vert N\right\rangle _{b}\pm \left\vert N\right\rangle
_{a}\left\vert \alpha \right\rangle _{b}$, and (v) entangled \textit{%
particle-like} NOON states $\left\vert N\right\rangle _{a}\left\vert
0\right\rangle _{b}\pm \left\vert 0\right\rangle _{a}\left\vert
N\right\rangle _{b}$\ ($N$ is a positive integer). The first two have
applications in quantum teleportation [34,35] and quantum key distribution
[36], while the last two have applications in quantum metrology [37,38] and
precision measurement [39]. The third may have potential applications
because quantum information with \textit{cat-state }encoding qubits is
recently attracting considerable attention [40]. Moreover, our method can be
used to generate a set of three-oscillator entangled states, e.g.,
wave-wave-particle-like entangled states $\left\vert \alpha \right\rangle
_{a}\left\vert -\alpha \right\rangle _{b}\left\vert 1\right\rangle _{c}\pm
\left\vert -\alpha \right\rangle _{a}\left\vert \alpha \right\rangle
_{b}\left\vert 0\right\rangle _{c},$ $\left\vert \xi \right\rangle
_{a}\left\vert -\xi \right\rangle _{b}\left\vert 1\right\rangle _{c}\pm
\left\vert -\xi \right\rangle _{a}\left\vert \xi \right\rangle
_{b}\left\vert 0\right\rangle _{c},$ and $\left\vert \mathrm{cat}%
\right\rangle _{a}\left\vert \overline{\mathrm{cat}}\right\rangle
_{b}\left\vert 1\right\rangle _{c}\pm \left\vert \overline{\mathrm{cat}}%
\right\rangle _{a}\left\vert \mathrm{cat}\right\rangle _{b}\left\vert
0\right\rangle _{c}$; and particle-like entangled states $\left\vert
N\right\rangle _{a}\left\vert 0\right\rangle _{b}\left\vert 1\right\rangle
_{c}\pm \left\vert 0\right\rangle _{a}\left\vert N\right\rangle
_{b}\left\vert 0\right\rangle _{c}.$ These types of entangled states may
have applications in quantum crytography [41], quantum secret sharing [42],
and controlled quantum teleportation [43]. Furthermore, the protocol can be
used to generate many other different types of two-oscillator or
three-oscillator (known or unknown) entangled states that are not mentioned
above.

As shown below, the entanglement generation operates essentially via the
quantum state swapping conditioned on the state of the coupler. Namely, when
the coupler is in the state $\left\vert g^{\prime }\right\rangle ,$ the
two-oscillator initial state $\left\vert \varphi \right\rangle
_{a}\left\vert \phi \right\rangle _{b}$ remains unchanged; however, when the
coupler is in the state $\left\vert g\right\rangle ,$ the two-oscillator
initial state $\left\vert \varphi \right\rangle _{a}\left\vert \phi
\right\rangle _{b}$ changes to $\left\vert \phi \right\rangle _{a}\left\vert
\varphi \right\rangle _{b}$ via the state swapping $\left\vert \varphi
\right\rangle \leftrightarrow \left\vert \phi \right\rangle .$ Hence, the
physical mechanism used for the entanglement creation here is quite
different from those based on state synthesis algorithms [44-48] which
require applying a sequence of operations in order to prepare the desired
states. The number of operations, required by state-synthesis algorithms for
preparing the target states $\left\vert \Psi \right\rangle _{\text{target}%
}=\sum_{m,n}C_{mn}\left\vert m,n\right\rangle $, increases drastically with
the dimensionality of the subspace of the Fock-state space in which the
target states are embedded [44-48].

\textit{Hamiltonian and intuition.} Two oscillators $a$ and $b$ are coupled
to a coupler with an energy level $\left\vert g\right\rangle $. The
Hamiltonian considered here is given by (assuming $\hbar =1$)
\begin{equation}
H=\omega \left( \hat{a}^{\dagger }\hat{a}+\hat{b}^{\dagger }\hat{b}\right)
\left\vert g\right\rangle \left\langle g\right\vert +\lambda \left( \hat{a}%
^{\dagger }\hat{b}+\hat{a}\hat{b}^{\dagger }\right) \left\vert
g\right\rangle \left\langle g\right\vert ,
\end{equation}%
where $a$ ($b$) is the photon annihilation operator of oscillator $a$ ($b$),
$\left\vert \omega \right\vert $ ($\omega $ being either positive or
negative)\ is the frequency or frequency shift of both oscillators, and $%
\left\vert \lambda \right\vert $ ($\lambda $ being either positive or
negative) is the coupling strength between the two oscillators. The second
term $\lambda \left( \hat{a}^{\dagger }\hat{b}+\hat{a}\hat{b}^{\dagger
}\right) \left\vert g\right\rangle \left\langle g\right\vert $ represents
the interaction between the two oscillators when the coupler is in the state
$\left\vert g\right\rangle .$ After some interaction time, this term results
in the exchange of the states of the two oscillators when the coupler is in
the state $\left\vert g\right\rangle .$ However, the two-oscillator state
exchange is imperfect without including the first term $\omega \left( \hat{a}%
^{\dagger }\hat{a}+\hat{b}^{\dagger }\hat{b}\right) \left\vert
g\right\rangle \left\langle g\right\vert ,$ because the state exchange
resulting from the second term $\lambda \left( \hat{a}^{\dagger }\hat{b}+%
\hat{a}\hat{b}^{\dagger }\right) \left\vert g\right\rangle \left\langle
g\right\vert $ comes with inevitable photon-number-dependent phase errors.
For instance, the state $\left\vert \varphi \right\rangle
=\sum\limits_{n=0}^{\infty }c_{n}\left\vert n\right\rangle $ of oscillator $a
$ (with $\left\vert n\right\rangle $ being the $n$-photon Fock state) is
transferred onto oscillator $b$ initially in a vacuum state by an error
state $\left\vert \varphi \right\rangle _{\mathrm{er}}=\sum\limits_{n=0}^{%
\infty }c_{n}e^{i\phi _{n}}\left\vert n\right\rangle $ (see the discussion
below).

Note that Eq.~(1) is different from the well-known Hamiltonian $\widetilde{H}%
=$ $\omega \left( \hat{a}^{\dagger }\hat{a}+\hat{b}^{\dagger }\hat{b}\right)
+\lambda \left( \hat{a}^{\dagger }\hat{b}+\hat{a}\hat{b}^{\dagger }\right) $
describing two single-mode interacting oscillators. This is because each
term in Eq.~(1) contains a coupler operator $\left\vert g\right\rangle
\left\langle g\right\vert ,$ which is however not involved in $\widetilde{H}.$

\textit{Entangling oscillators.} Suppose that oscillator $a$ is in an
arbitrary pure state $\left\vert \varphi \right\rangle _{a}$ and oscillator $%
b$ is in another arbitrary pure state $\left\vert \phi \right\rangle _{b}$.
Assume that a coupler is in a superposition state $\alpha \left\vert
g^{\prime }\right\rangle +\beta \left\vert g\right\rangle ,$ with $%
\left\vert \alpha \right\vert ^{2}+\left\vert \beta \right\vert ^{2}=1.$
Here, $\left\vert g^{\prime }\right\rangle $ is an excited state of the
coupler. Under the Hamiltonian in Eq.~(1), the initial state of the system $%
\left\vert \varphi \right\rangle _{a}\left\vert \phi \right\rangle
_{b}\left( \alpha \left\vert g^{\prime }\right\rangle +\beta \left\vert
g\right\rangle \right) $ evolves into
\begin{eqnarray}
&&e^{-iHt}\left\vert \varphi \right\rangle _{a}\left\vert \phi \right\rangle
_{b}\left( \alpha \left\vert g^{\prime }\right\rangle +\beta \left\vert
g\right\rangle \right)  \notag \\
&=&\alpha \left\vert \varphi \right\rangle _{a}\left\vert \phi \right\rangle
_{b}\left\vert g^{\prime }\right\rangle +\beta \left( e^{-iH_{e}t}\left\vert
\varphi \right\rangle _{a}\left\vert \phi \right\rangle _{b}\right) \otimes
\left\vert g\right\rangle ,
\end{eqnarray}%
where we have used $\left\langle g\right. \left\vert g^{\prime
}\right\rangle =0.$ Here, $H_{e}=H_{0}+H_{I}$ with $H_{0}=\omega \left( \hat{%
a}^{\dagger }\hat{a}+\hat{b}^{\dagger }\hat{b}\right) $ and $H_{I}=\lambda
\left( \hat{a}^{\dagger }\hat{b}+\hat{a}\hat{b}^{\dagger }\right) .$ $H_{e}$
describes the dynamics of the oscillators, which arises from Eq.~(1) when
the coupler is in the state $\left\vert g\right\rangle .$ Because of $\left[
H_{0},H_{I}\right] =0,$ the oscillator state $e^{-iH_{e}t}\left\vert \varphi
\right\rangle _{a}\left\vert \phi \right\rangle _{b}$ of Eq.~(2) can be
written as
\begin{equation}
e^{-iH_{e}t}\left\vert \varphi \right\rangle _{a}\left\vert \phi
\right\rangle _{b}=U_{2}U_{1}\left\vert \varphi \right\rangle _{a}\left\vert
\phi \right\rangle _{b},
\end{equation}%
with $U_{1}=e^{-iH_{I}t}$ and $U_{2}=e^{-iH_{0}t}.$

$U_{1}$ leads to the transformations $U_{1}\hat{a}^{\dagger }U_{1}^{+}=\cos
(\lambda t)\hat{a}^{\dagger }-i\sin (\lambda t)\hat{b}^{\dagger }$ and $U_{1}%
\hat{b}^{\dagger }U_{1}^{+}=\cos (\lambda t)\hat{b}^{\dagger }-i\sin
(\lambda t)\hat{a}^{\dagger }.$ For $\left\vert \lambda \right\vert t=\left(
2m+1/2\right) \pi $ ($m$ is an integer), one has $U_{1}\left( \hat{a}%
^{\dagger }\right) ^{n}U_{1}^{+}=\left( \mp i\hat{b}^{\dagger }\right) ^{n}$
and $U_{1}\left( \hat{b}^{\dagger }\right) ^{n}U_{1}^{+}=\left( \mp i\hat{a}%
^{\dagger }\right) ^{n},$ which will be applied in derivation of Eq.~(5)
below. Here and below, the sign \textquotedblleft $-$\textquotedblright\
corresponds to $\lambda >0$ while \textquotedblleft $+$\textquotedblright\
corresponds to $\lambda <0.$ The arbitrary pure states $\left\vert \varphi
\right\rangle _{a}$ and $\left\vert \phi \right\rangle _{b}$ can be
expressed as
\begin{equation}
\left\vert \varphi \right\rangle _{a}=\sum_{n=0}^{\infty }c_{n}\left\vert
n\right\rangle _{a},\text{ \ }\left\vert \phi \right\rangle
_{b}=\sum_{m=0}^{\infty }d_{m}\left\vert m\right\rangle _{b},
\end{equation}%
where $c_{n}$ and $d_{m}$ are normalized coefficients, $\left\vert
n\right\rangle _{a}=\frac{\left( \hat{a}^{\dagger }\right) ^{n}}{\sqrt{n!}}%
\left\vert 0\right\rangle _{a}$ ($\left\vert m\right\rangle _{b}=\frac{%
\left( \hat{b}^{\dagger }\right) ^{m}}{\sqrt{m!}}\left\vert 0\right\rangle
_{b}$) representing the $n$-photon ($m$-photon) Fock state of oscillator $a$
($b$).

By performing a unitary transformation $U_{1},$ after $t=\pi /\left(
2\left\vert \lambda \right\vert \right) ,$ the state $\left\vert \varphi
\right\rangle _{a}\left\vert \phi \right\rangle _{b}$ evolves into
\begin{eqnarray}
&&U_{1}\left\vert \varphi \right\rangle _{a}\left\vert \phi \right\rangle
_{b}  \notag \\
&=&\sum\limits_{n=0}^{\infty }\sum\limits_{m=0}^{\infty }\frac{c_{n}d_{m}}{%
\sqrt{n!m!}}\left[ U_{1}\left( \hat{a}^{\dagger }\right) ^{n}U_{1}^{+}\right]
\left[ U_{1}\left( \hat{b}^{\dagger }\right) ^{m}U_{1}^{+}\right]
U_{1}\left\vert 0\right\rangle _{a}\left\vert 0\right\rangle _{b}  \notag \\
&=&\sum_{n=0}^{\infty }c_{n}\left( \mp i\right) ^{n}\frac{\left( \hat{b}%
^{\dagger }\right) ^{n}}{\sqrt{n!}}\left\vert 0\right\rangle _{a}\times
\sum_{m=0}^{\infty }d_{m}\left( \mp i\right) ^{m}\frac{\left( \hat{a}%
^{\dagger }\right) ^{m}}{\sqrt{m!}}\left\vert 0\right\rangle _{b}  \notag \\
&=&\sum_{n=0}^{\infty }c_{n}e^{\mp in\pi /2}\left\vert n\right\rangle
_{b}\otimes \sum_{m=0}^{\infty }d_{m}e^{\mp im\pi /2}\left\vert
m\right\rangle _{a},
\end{eqnarray}%
where the positions of $\left\vert 0\right\rangle _{a}$ and $\left\vert
0\right\rangle _{b}$ in line 3 are exchanged in the last line and $%
U_{1}\left\vert 0\right\rangle _{a}\left\vert 0\right\rangle _{b}=\left\vert
0\right\rangle _{a}\left\vert 0\right\rangle _{b}$ is applied. The first
(second)\ part of the product in the last line represents the state of
oscillator $b$ ($a$). Comparing the last line with the original states $%
\left\vert \varphi \right\rangle _{a}$ and $\left\vert \phi \right\rangle
_{b}$ given in Eq.~(4), one can see that the two oscillators exchange their
states while accumulating photon-number-dependent phase errors $e^{\mp in\pi
/2}$ and\ $e^{\mp im\pi /2},$ respectively.

By performing a unitary transformation $U_{2}$ with $t=\pi /\left(
2\left\vert \lambda \right\vert \right) $ and setting $\mp \pi /2-\omega
t=2k\pi $ ($k$ is an integer), the state (5) becomes
\begin{eqnarray}
&&U_{2}\left( U_{1}\left\vert \varphi \right\rangle _{a}\left\vert \phi
\right\rangle _{b}\right)  \notag \\
&=&\sum_{n=0}^{\infty }c_{n}e^{in\left( \mp \pi /2-\omega t\right)
}\left\vert n\right\rangle _{b}\otimes \sum_{m=0}^{\infty }d_{m}e^{im\left(
\mp \pi /2-\omega t\right) }\left\vert m\right\rangle _{a}  \notag \\
&=&\sum_{n=0}^{\infty }c_{n}\left\vert n\right\rangle _{b}\otimes
\sum_{m=0}^{\infty }d_{m}\left\vert m\right\rangle _{a}=\left\vert \varphi
\right\rangle _{b}\left\vert \phi \right\rangle _{a},
\end{eqnarray}%
where $\left\vert \varphi \right\rangle _{b}$ ($\left\vert \phi
\right\rangle _{a}$) takes the same form of the state $\left\vert \varphi
\right\rangle _{a}$ ($\left\vert \phi \right\rangle _{b}$) with the
subscript $a$ ($b$) replaced by $b$ ($a$). Combining Eqs.~(3) and (6), one
finds that the state (2) would be
\begin{equation}
\alpha \left\vert \varphi \right\rangle _{a}\left\vert \phi \right\rangle
_{b}\left\vert g^{\prime }\right\rangle +\beta \left\vert \phi \right\rangle
_{a}\left\vert \varphi \right\rangle _{b}\left\vert g\right\rangle .
\end{equation}%
Now apply a classical pulse to the coupler, resulting in $\left\vert
g^{\prime }\right\rangle \rightarrow $ $\left( \left\vert g\right\rangle
+\left\vert g^{\prime }\right\rangle \right) /\sqrt{2}$ and $\left\vert
g\right\rangle \rightarrow \left( \left\vert g\right\rangle -\left\vert
g^{\prime }\right\rangle \right) /\sqrt{2}.$ Thus, the state (7) becomes
\begin{equation}
\frac{1}{\sqrt{2}}\left( \left\vert \psi ^{+}\right\rangle \otimes
\left\vert g\right\rangle +\left\vert \psi ^{-}\right\rangle \otimes
\left\vert g^{\prime }\right\rangle \right) ,
\end{equation}%
with
\begin{equation}
\left\vert \psi ^{\pm }\right\rangle =\alpha \left\vert \varphi
\right\rangle _{a}\left\vert \phi \right\rangle _{b}\pm \beta \left\vert
\phi \right\rangle _{a}\left\vert \varphi \right\rangle _{b}.
\end{equation}%
Eq.~(8) shows that when the coupler is measured in the state $\left\vert
g\right\rangle $ ($\left\vert g^{\prime }\right\rangle $)$,$ the two
oscillators are prepared in an entangled state $\left\vert \psi
\right\rangle ^{+}$ ($\left\vert \psi ^{-}\right\rangle $), for which the
degree of entanglement can be adjusted by varying $\alpha $ and $\beta $
during the preparation of the initial state of the coupler.

It is straightforward to show that the state (7) can be transformed to a
three-oscillator entangled state
\begin{equation}
\alpha \left\vert \varphi \right\rangle _{a}\left\vert \phi \right\rangle
_{b}\left\vert 1\right\rangle _{c}+\beta \left\vert \phi \right\rangle
_{a}\left\vert \varphi \right\rangle _{b}\left\vert 0\right\rangle _{c},
\end{equation}%
by performing local operations on the coupler and a third oscillator $c$
initially in the vacuum state. For instance, this transformation from the
state (7) to the state (10) can be achieved by tuning the frequency of
oscillator $c$ on resonance with the $\left\vert g\right\rangle
\leftrightarrow \left\vert g^{\prime }\right\rangle $ transition or vice
versa, to have a single photon emitted into oscillator $c$ when the coupler
is in the excited state $\left\vert g^{\prime }\right\rangle $.

\textit{Hamiltonian construction.} The four levels of the coupler are
denoted as $\left\vert g\right\rangle ,$ $\left\vert g^{\prime
}\right\rangle ,$ $\left\vert e\right\rangle ,$ and $\left\vert
f\right\rangle $ [Fig.~1(a)]. The level $\left\vert g^{\prime }\right\rangle
$ can remain unaffected, for example, by having the transition between $%
\left\vert g^{\prime }\right\rangle $ and any other level highly detuned
from the frequencies of the two oscillators and the classical pulse.
Oscillator $a$ ($b$) is coupled to the $\left\vert g\right\rangle $ $%
\leftrightarrow $ $\left\vert f\right\rangle $ ($\left\vert g\right\rangle $
$\leftrightarrow $ $\left\vert e\right\rangle $)\ transition with coupling
strength $g_{a}$ ($g_{b}$) and detuning $\Delta _{a}=\omega _{fg}-\omega
_{a} $ ($\delta _{b}=\omega _{eg}-\omega _{b}$) [Fig.~1(a)]. Here, $\omega
_{fg}$ ($\omega _{eg}$) is the $\left\vert g\right\rangle \leftrightarrow $ $%
\left\vert f\right\rangle $ ($\left\vert g\right\rangle $ $\leftrightarrow $
$\left\vert e\right\rangle $) transition frequency and $\omega _{a}$ ($%
\omega _{b}$) is the frequency of oscillator $a$ ($b$). A classical pulse of
frequency $\omega _{p}$ is coupled to the $\left\vert e\right\rangle $ $%
\leftrightarrow $ $\left\vert f\right\rangle $ transition with detunings $%
\Delta =\omega _{fe}-\omega _{p}$ [Fig.~1(a)]$.$ In the interaction picture
under the free Hamiltonian $H_{\mathrm{field}}+H_{\mathrm{atom}}$ with\ $H_{%
\mathrm{field}}=\omega _{a}\hat{a}^{\dagger }\hat{a}+\omega _{b}\hat{b}%
^{\dagger }\hat{b}$, the Hamiltonian is given by
\begin{eqnarray}
H &=&\left( g_{a}e^{i\Delta _{a}t}\hat{a}\sigma _{fg}^{+}+g_{b}e^{i\delta
_{b}t}\hat{b}\sigma _{eg}^{+}+\text{H.c.}\right)  \notag \\
&&\ +\left( \Omega e^{i\Delta t}\sigma _{fe}^{+}+\text{H.c.}\right) ,
\end{eqnarray}%
where $\sigma _{fg}^{+}=\left\vert f\right\rangle \left\langle g\right\vert $%
, $\sigma _{fe}^{+}=\left\vert f\right\rangle \left\langle e\right\vert ,$ $%
\Omega $ is the Rabi frequency of the classical pulse, and $\hat{a}$ ($\hat{b%
}$) is the photon annihilation operator of oscillator $a$ ($b$).

Under large-detuning conditions and when the levels $\left\vert
e\right\rangle $ and $\left\vert f\right\rangle $ are not occupied, the
Hamiltonian of Eq.~(11) can be expressed as the following effective
Hamiltonian (see Supplemental Material)
\begin{eqnarray}
H_{\mathrm{eff}} &=&-\left( g_{a}^{2}/\Delta _{a}+\widetilde{g}%
_{a}^{2}/\delta \right) \hat{a}^{\dagger }\hat{a}\left\vert g\right\rangle
\left\langle g\right\vert -g_{b}^{2}/\delta \hat{b}^{\dagger }\hat{b}%
\left\vert g\right\rangle \left\langle g\right\vert   \notag \\
&&+\lambda \left( \hat{a}\hat{b}^{\dagger }+\hat{a}^{\dagger }\hat{b}\right)
\left\vert g\right\rangle \left\langle g\right\vert ,
\end{eqnarray}%
where $\tilde{g}_{a}=g_{a}\Omega (\Delta _{a}^{-1}+\Delta ^{-1})/2,$ $\delta
_{a}=\Delta _{a}-\Delta ,$ and $\lambda =\widetilde{g}_{a}g_{b}/\delta >0.$
In Eq.~(12), we have set $\delta _{a}=\delta _{b}\equiv \delta >0,$ i.e., $%
\omega _{p}=\omega _{a}-\omega _{b},$ which can be readily achieved by
adjusting the pulse frequency $\omega _{p}$. By setting
\begin{equation}
\frac{g_{a}^{2}}{\Delta _{a}}+\frac{g_{a}^{2}\Omega ^{2}}{4\delta }\left(
\Delta _{a}^{-1}+\Delta ^{-1}\right) ^{2}=\frac{g_{b}^{2}}{\delta }=-\omega ,
\end{equation}%
(e.g., by adjusting the pulse Rabi frequency $\Omega $), one sees that Eq.~(12) 
takes the same form as the Hamiltonian (1). Based on Eq.~(13) and
setting $\mp \pi /2-\omega t=2k\pi $, we can obtain the following
relationship between the various parameters
\begin{eqnarray}
g_{b} &=&\frac{\left\vert 4k\pm 1\right\vert }{2\sqrt{2k\left( 2k\pm
1\right) \Delta _{a}/\delta }}g_{a},  \notag \\
\Omega  &=&\frac{\Delta \Delta _{a}}{\Delta +\Delta _{a}}\sqrt{\delta /\left[
2k\left( 2k\pm 1\right) \Delta _{a}\right] },
\end{eqnarray}%
which shows that the pulse Rabi frequency $\Omega $ is independent of the
coupling strengths $g_{a}$ and $g_{b}.$

Note that the four-level structure in Fig.~1(a) is widely available in
natural or artificial atoms such as quantum dots, NV centers, and various
superconducting devices [49]. Thus, the Hamiltonian (1) can be realized with
a variety of physical systems. As shown above, the Hamiltonian (12), i.e.,
Eq.~(1), was constructed based on the Raman transition induced by the
field-pulse cooperation. Note that it is possible to construct the proposed
Hamiltonian (1) based on other physical mechanisms.

\begin{figure}[tbp]
\begin{center}
\includegraphics[bb=14 13 601 301, width=8.5 cm, clip]{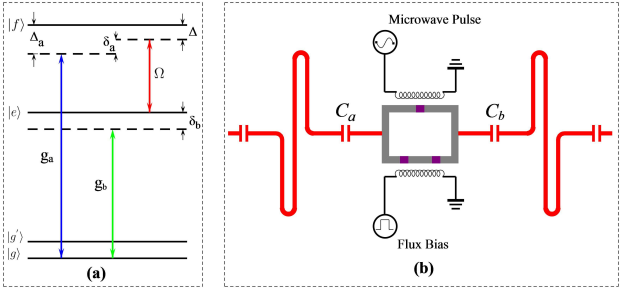} \vspace*{%
-0.08in}
\end{center}
\caption{(color online). (a). Illustration of the coupler interacting with
two oscillators and a classical pulse. Here, $\protect\delta_a=\protect\omega%
_p+\protect\omega_{eg}-\protect\omega_a=\protect\delta_b$, which can be
readily met by adjusting the pulse frequency $\protect\omega_p$. (b). Set-up
of two cavities coupled to a flux device via a capacitor $C_a$ or $C_b$.}
\label{fig:1}
\end{figure}

\textit{Circuit-QED Implementation.} Circuit QED with resonators and
superconducting qubits is one of the most promising candidates for quantum
information processing (for reviews, see [50-53]). We now consider a setup
consisting of two microwave resonators coupled via a superconducting artificial atom
[Fig.~1(b)]. Each resonator here is a 1D transmission line resonator
(TLR). The four levels of the coupler are illustrated in Fig.~1(a). The
pulse- or resonator-induced unwanted transitions between irrelevant levels
are assumed to be negligibly small. This can be achieved by a prior design
of the coupler with a strong anharmonicity (e.g., a
superconducting flux device). Alternatively, this condition can be satisfied
by adjusting the coupler level spacings or the resonator frequencies. In
practice, level spacings of superconducting devices can be rapidly adjusted
within a few nanoseconds (e.g., see [54] and references therein) and, to a
lesser extent, frequencies of the resonators can be fast tuned in 1-3 ns
[55,56]. When the inter-resonator crosstalk is taken into account, the
Hamiltonian (11) becomes $H^{\prime }=H+\varepsilon ,$ where $\varepsilon $
describes the unwanted inter-resonator crosstalk, given by $\varepsilon
=g_{ab}e^{i\Delta _{ab}t}\hat{a}^{\dagger }\hat{b}+h.c.,$ with the
two-resonator coupling strength $g_{ab}$ and the resonator frequency
detuning $\Delta _{ab}=\omega _{a}-\omega _{b}.$ Here, $\omega _{a}$ ($%
\omega _{b}$) is the frequency of resonator $a$ ($b$).

\begin{figure}[tbp]
\begin{center}
\includegraphics[bb=0 69 865 659, width=8.0 cm, clip]{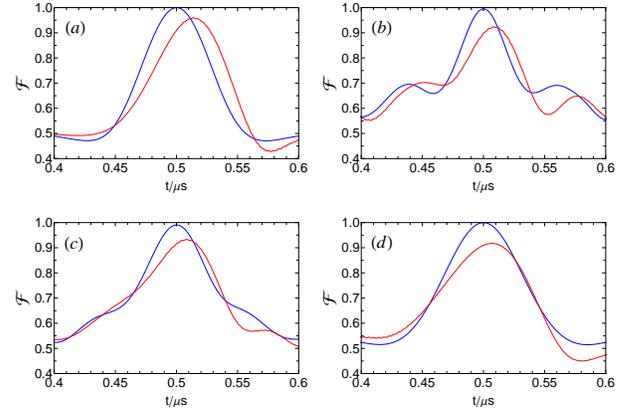} \vspace*{%
-0.08in}
\end{center}
\caption{(color online). Fidelities versus the operation time $t$. (a), (b),
(c), and (d) are for entangled coherent states, entangled squeezed states,
entangled coherent-squeezed states, and entangled cat states, respectively.
Blue curves were based on the effective Hamiltonian (12) without
considering decoherence; while red curves were based on the master equation (15) by
taking decoherence into consideration.}
\label{fig:2}
\end{figure}

The fidelity of the operation is given by $\mathcal{F}=\sqrt{\left\langle
\psi _{id}\right\vert \rho \left\vert \psi _{id}\right\rangle },$ where $%
\left\vert \psi _{id}\right\rangle $ is the ideal state given in Eq.~(7),
while $\rho $ is the final density operator of the whole system after the
operation is performed in a realistic system. As an example, we consider $%
\alpha =\beta =1/\sqrt{2}$.

By solving the master equation and choosing the system parameters
appropriately (see Supplemental Material), the simulated fidelity $%
\mathcal{F}$ versus the operation time $t$ are shown in Fig.~2 for $\eta
=\Delta _{a}/g_{a}=25,$ $k=1$ and $\alpha =$ $\xi =1$, where $\left\vert \pm
\xi \right\rangle $ are squeezed vacuum states. One can see that for $t\sim
0.5$ $\mu $s, a high fidelity can be obtained: (i) $\mathcal{F}\simeq 0.959$
for the entangled coherent states $\frac{1}{\sqrt{2}}\left( \left\vert
\alpha \right\rangle _{a}\left\vert -\alpha \right\rangle _{b}\pm \left\vert
-\alpha \right\rangle _{a}\left\vert \alpha \right\rangle _{b}\right) $
[Fig.~2(a)]; (ii) $\mathcal{F}\simeq 0.912$ for the entangled squeezed
states $\frac{1}{\sqrt{2}}\left( \left\vert \xi \right\rangle _{a}\left\vert
-\xi \right\rangle _{b}\pm \left\vert -\xi \right\rangle _{a}\left\vert \xi
\right\rangle _{b}\right) $ [Fig.~2(b)]; (iii) $\mathcal{F}\simeq 0.929$ for
the entangled coherent-squeezed states $\frac{1}{\sqrt{2}}\left( \left\vert
\alpha \right\rangle _{a}\left\vert \xi \right\rangle _{b}\pm \left\vert \xi
\right\rangle _{a}\left\vert \alpha \right\rangle _{b}\right) $ [Fig.~2(c)];
and (iv) $\mathcal{F}\simeq 0.918$ for the entangled cat states $\frac{1}{%
\sqrt{2}}\left( \left\vert \mathrm{cat}\right\rangle _{a}\left\vert
\overline{\mathrm{cat}}\right\rangle _{b}\pm \left\vert \overline{\mathrm{cat%
}}\right\rangle _{a}\left\vert \mathrm{cat}\right\rangle _{b}\right) $.

For $\eta =25$, we have $g_{a}/2\pi \sim $ $60$ MHz, $g_{b}/2\pi \sim 25$
MHz, and $\Omega /2\pi \sim 114$ MHz, which are available in experiments
[57,58]. The frequency of a circuit resonator is typically a few GHz.
For the sake of concreteness, consider $\omega _{a}/\left( 2\pi \right) \sim 7.5$ GHz and $\omega
_{b}/\left( 2\pi \right) \sim 4.5$ GHz. For the values of $\kappa _{a}^{-1}$
and $\kappa _{b}^{-1}$ used in the numerical simulation, the required
quality factors for the two resonators are $Q_{a}\sim 9.4\times 10^{5}$ and $%
Q_{b}\sim 5.6\times 10^{5},$ readily available in experiments [59,60]. The
analysis here demonstrates that by applying the proposed protocol, the
high-fidelity generation of entanglement between asymmetric states of two
oscillators is feasible with current circuit QED technology. Finally, we
remark that the fidelity obtained above was calculated without considering
the initial state preparation and measurement errors, which however could be
negligible due to progress in accurate preparation and measurement of the
states of superconducting artificial atoms [61].

Finally, it is interesting to note that based on the Hamiltonian (1), when
the coupler is in the state $\left\vert g\right\rangle $, a SWAP
gate of two {\it discrete-variable} qubits or two {\it continuous-variable} qubits,
defined by $\left\vert \varphi \right\rangle _{a}\left\vert \varphi
\right\rangle _{b}\rightarrow $ $\left\vert \varphi \right\rangle
_{a}\left\vert \varphi \right\rangle _{b},$ $\left\vert \varphi
\right\rangle _{a}\left\vert \phi \right\rangle _{b}\rightarrow $ $%
\left\vert \phi \right\rangle _{a}\left\vert \varphi \right\rangle _{b},$ $%
\left\vert \phi \right\rangle _{a}\left\vert \varphi \right\rangle
_{b}\rightarrow $ $\left\vert \varphi \right\rangle _{a}\left\vert \phi
\right\rangle _{b},$ and $\left\vert \phi \right\rangle _{a}\left\vert \phi
\right\rangle _{b}\rightarrow $ $\left\vert \phi \right\rangle
_{a}\left\vert \phi \right\rangle _{b},$ can be realized without
measurement. Here, a qubit is encoded by the two states $\left\vert \varphi
\right\rangle $ and $\left\vert \phi \right\rangle $ of each oscillator. For
$\left\vert \varphi \right\rangle =$ $\left\vert \mathrm{cat}\right\rangle $
and $\left\vert \phi \right\rangle =\left\vert \overline{\mathrm{cat}}%
\right\rangle ,$ the two-qubit SWAP gate is implemented with \textit{%
cat-state} encoding qubits which attract increasing attention recently [40].

\textit{Acknowledgments.} C.P.Y. and Q.P.S. were supported by the National
Natural Science Foundation of China under Grant Nos.~[11074062,~11374083,~
11504075] and the Zhejiang Natural Science Foundation under Grant
No.~LZ13A040002. S.B.Z. was supported by the Major State Basic Research
Development Program of China under Grant No.~2012CB921601. F.N. was
supported by the RIKEN iTHES Project, the MURI Center for Dynamic
Magneto-Optics via the AFOSR award number FA9550-14-1-0040, a Grant-in-Aid
for Scientific Research (A), and a grant from the John Templeton Foundation.
S.H. was supported by the NSF (Grant No. PHY-1314861). This work was also
supported by the funds of Hangzhou City for supporting the Hangzhou-City
Quantum Information and Quantum Optics Innovation Research Team.

\begin{center}
\textbf{\textbf{Supplementary material for entangling two oscillators with arbitrary asymmetric initial states}}
\end{center}

\begin{center}
\textbf{\textbf{Derivation of an effective Hamiltonian}}
\end{center}

Let us start with the original Hamiltonian given in Eq.~(11), i.e.,

\begin{eqnarray}
H &=&g_{a}(\hat{a}\sigma _{fg}^{+}e^{i\Delta _{a}t}+h.c.)+g_{b}(\hat{b}%
\sigma _{eg}^{+}e^{i\delta _{b}t}+h.c.)  \notag \\
&&+\Omega (e^{i\Delta t}\sigma _{fe}^{+}+h.c.),  \label{S1}
\end{eqnarray}
where $\sigma _{eg}^{+}=|e\rangle \langle g|$ and $\sigma
_{fg}^{+}=|f\rangle \langle g|$, $\Omega $ is the Rabi frequency of the
pulse, and $\hat{a}$ ($\hat{b}$) is the photon annihilation operator for
quantum oscillator $a$ ($b$).

Under the large-detuning conditions $\Delta _{a}\gg g_{a}$ and $\Omega \gg
\Delta ,$ there is no energy exchange between oscillator $a$ and the
coupler, as well as between the pulse and the coupler [Fig. 1(a)]. In
addition, under the conditions $\Delta _{a}-\delta _{b}\gg g_{a}g_{b}\left(
\Delta _{a}^{-1}+\delta _{b}^{-1}\right) /2$ and $\Delta -\delta _{b}\gg
\Omega g_{b}\left( \Delta ^{-1}+\delta _{b}^{-1}\right) /2,$ there is no
interaction between oscillator $b$ and either of oscillator $a$ and the
pulse [Fig. 1(a)]. In this case, the effective Hamiltonian can be expressed
as [1]
\begin{eqnarray}
H_{\mathrm{eff}} &=&\frac{g_{a}^{2}}{\Delta _{a}}[~|f\rangle \langle f|+\hat{%
a}^{\dagger }\hat{a}(|f\rangle \langle f|-|g\rangle \langle g|)~]  \notag \\
&&+\frac{\Omega ^{2}}{\Delta }(|f\rangle \langle f|-|e\rangle \langle e|)
\notag \\
&&-\tilde{g}_{a}(\hat{a}\sigma _{eg}^{+}e^{i\delta _{a}t}+h.c.)  \notag \\
&&+g_{b}(\hat{b}\sigma _{eg}^{+}e^{i\delta _{b}t}+h.c.),  \label{S2}
\end{eqnarray}
where $\tilde{g}_{a}=g_{a}\Omega (\Delta _{a}^{-1}+\Delta ^{-1})/2$ and $%
\delta _{a}=\Delta _{a}-\Delta $. Under the large-detuning conditions $%
\delta _{a}\gg \{\tilde{g}_{a},g_{a}^{2}/\Delta _{a},\Omega ^{2}/\Delta \}$
and $\delta _{b}\gg \{g_{b},g_{a}^{2}/\Delta _{a},\Omega ^{2}/\Delta \}$,
the effective Hamiltonian $H_{\mathrm{eff}}$ becomes [1]
\begin{eqnarray}
H_{\mathrm{eff}} &=&\frac{\tilde{g}_{a}^{2}}{\delta }[|e\rangle \langle e|+%
\hat{a}^{\dagger }\hat{a}(|e\rangle \langle e|-|g\rangle \langle g|)]  \notag
\\
&&+\frac{g_{b}^{2}}{\delta _{b}}[|e\rangle \langle e|+\hat{b}^{\dagger }\hat{%
b}(|e\rangle \langle e|-|g\rangle \langle g|)]  \notag \\
&&+\frac{g_{a}^{2}}{\Delta _{a}}[|f\rangle \langle f|+\hat{a}^{\dagger }\hat{%
a}(|f\rangle \langle f|-|g\rangle \langle g|)]  \notag \\
&&+\frac{\Omega ^{2}}{\Delta }(|f\rangle \langle f|-|e\rangle \langle e|)
\notag \\
&&-\frac{\tilde{g}_{a}g_{b}}{2}(\frac{1}{\delta _{a}}+\frac{1}{\delta _{b}}%
)\times   \notag \\
&&\lbrack (\hat{a}\hat{b}^{\dagger }|e\rangle \langle e|-~\hat{a}^{\dagger }%
\hat{b}|g\rangle \langle g|)e^{i(\delta _{a}-\delta _{b})t}+h.c.].
\label{S3}
\end{eqnarray}

When the levels $|e\rangle $ and $|f\rangle $ are not occupied, the
effective Hamiltonian $H_{\mathrm{eff}}$ reduces to
\begin{eqnarray}
H_{eff} &=&-(\frac{g_{a}^{2}}{\Delta _{a}}+\frac{\tilde{g}_{a}^{2}}{\delta })%
\hat{a}^{\dagger }\hat{a}|g\rangle \langle g|-\frac{g_{b}^{2}}{\delta }\hat{b%
}^{\dagger }\hat{b}|g\rangle \langle g|  \notag \\
&&+\lambda (\hat{a}\hat{b}^{\dagger }+\hat{a}^{\dagger }\hat{b})|g\rangle
\langle g|,  \label{S4}
\end{eqnarray}%
where $\lambda =\tilde{g}_{a}g_{b}/\delta $ and we have set $\delta
_{a}=\delta _{b}=\delta $.

\begin{center}
\textbf{\textbf{Master equation and parameters used in numerical simulations}}
\end{center}

After taking dissipation and dephasing into account, the system dynamics is
determined by the master equation
\begin{eqnarray}
\frac{d\rho }{dt} &=&-i\left[ H^{\prime },\rho \right] +\kappa _{a}\mathcal{L%
}\left[ \hat{a}\right] +\kappa _{b}\mathcal{L}\left[ \hat{b}\right]   \notag
\\
&&+\sum\limits_{j=g,g^{\prime },e}\gamma _{fj}\mathcal{L}\left[ \sigma
_{fj}^{-}\right] +\sum\limits_{k=g,g^{\prime }}\gamma _{ek}\mathcal{L}\left[
\sigma _{ek}^{-}\right] +\gamma _{g^{\prime }g}\mathcal{L}\left[ \sigma
_{g^{\prime }g}^{-}\right]   \notag \\
&&+\sum\limits_{j=g^{\prime },e,f}\gamma _{\varphi ,l}\left( \sigma
_{ll}\rho \sigma _{ll}-\sigma _{ll}\rho /2-\rho \sigma _{ll}/2\right) ,
\label{S5}
\end{eqnarray}%
where $\mathcal{L}\left[ \Lambda \right] =\Lambda \rho \Lambda ^{+}-\Lambda
^{+}\Lambda \rho /2-\rho \Lambda ^{+}\Lambda /2$ (with $\Lambda =\hat{a},%
\hat{b},\sigma _{g^{\prime }g}^{-},\sigma _{eg}^{-},\sigma _{eg^{\prime
}}^{-},\sigma _{fg}^{-},\sigma _{fg^{\prime }}^{-},\sigma _{fe}^{-}),$ $%
\sigma _{g^{\prime }g^{\prime }}=\left\vert g^{\prime }\right\rangle
\left\langle g^{\prime }\right\vert ,$ $\sigma _{ee}=\left\vert
e\right\rangle \left\langle e\right\vert ,$ and $\sigma _{ff}=\left\vert
f\right\rangle \left\langle f\right\vert $. In addition, $\kappa _{a}$ ($%
\kappa _{b}$) is the decay rate of resonator $a$ ($b$);\ $\gamma _{g^{\prime
}g}$, $\gamma _{eg}$, $\gamma _{eg^{\prime }}$, $\gamma _{fg}$, $\gamma
_{fg^{\prime }}$ and $\gamma _{fe}$ are the energy relaxation rates for $%
\left\vert g^{\prime }\right\rangle \rightarrow \left\vert g\right\rangle $,
$\left\vert e\right\rangle \rightarrow \left\vert g\right\rangle $, $%
\left\vert e\right\rangle \rightarrow \left\vert g^{\prime }\right\rangle $,
$\left\vert f\right\rangle \rightarrow \left\vert g\right\rangle $, $%
\left\vert f\right\rangle \rightarrow \left\vert g^{\prime }\right\rangle $,
and $\left\vert f\right\rangle \rightarrow \left\vert e\right\rangle $,
respectively; $\gamma _{\varphi ,g^{\prime }}$, $\gamma _{\varphi ,e}$, and $%
\gamma _{\varphi ,f}$ are the dephasing rates of the levels $\left\vert
g^{\prime }\right\rangle $, $\left\vert e\right\rangle $, and $\left\vert
f\right\rangle $.

The parameters used in the numerical simulation are: (i) $\Delta _{a}/2\pi
=1.5$ GHz, $\Delta /2\pi =1.25$ GHz; (ii) $\delta _{b}/2\pi =0.25$ GHz;
(iii) $\gamma _{\varphi ,g^{\prime }}^{-1}=\gamma _{\varphi ,e}^{-1}=\gamma
_{\varphi ,f}^{-1}=15$ $\mu $s; (iv) $\gamma _{g^{\prime }g}^{-1}=60$ $\mu $%
s, $\gamma _{eg^{\prime }}^{-1}=40$ $\mu $s, $\gamma _{fe}^{-1}=30$ $\mu $s,
$\gamma _{eg}^{-1}=$ $\gamma _{fg^{\prime }}^{-1}=\gamma _{fg}^{-1}=100$ $\mu $s [2];
and (v) $\kappa _{a}^{-1}=\kappa _{a}^{-1}=20$ $\mu $s. We choose $%
g_{12}=0.1\max \{g_{a},g_{b}\}$. Here we consider a rather conservative case
for both the inter-resonator crosstalk and the decoherence time of flux
qudits because the inter-resonator crosstalk strength can be smaller by at
least one order of magnitude [3] and decoherence time ranging from $70$ $\mu
$s to $1$ ms has been reported for a superconducting qudit [4-7].


\begin{thebibliography}{9}
\bibitem{s2} By designing the flux qudit, the $\left\vert e\right\rangle $ $%
\leftrightarrow \left\vert g\right\rangle $, $\left\vert f\right\rangle $ $%
\leftrightarrow \left\vert g^{\prime }\right\rangle $, and $\left\vert
f\right\rangle $ $\leftrightarrow \left\vert g\right\rangle $ dipole matrix
elements can be made much smaller than those of the $\left\vert g^{\prime
}\right\rangle \leftrightarrow \left\vert g\right\rangle ,$ $\left\vert
e\right\rangle \leftrightarrow \left\vert g^{\prime }\right\rangle $, and $%
\left\vert f\right\rangle \leftrightarrow \left\vert e\right\rangle $
transitions. Thus,  $\gamma _{eg}^{-1},\gamma _{fg^{\prime }}^{-1},\gamma
_{fg}^{-1}\gg \gamma _{g^{\prime }g}^{-1},\gamma _{eg^{\prime }}^{-1},\gamma
_{fe}^{-1}.$
\end{thebibliography}
\end{document}